\documentclass[journal]{IEEEtran}

\ifCLASSINFOpdf

\usepackage{graphicx}

\else

\fi

\begin{document}

\title{Resonant tunneling diode based on graphene/h-BN heterostructure}

\author{V.~Hung~Nguyen,~F.~Mazzamuto,~A.~Bournel,~and~P.~Dollfus
\thanks{V. Hung Nguyen is with the Institut d'Electronique Fondamentale, CNRS, Universit$\acute{e}$ Paris-Sud, UMR8622, 91405 Orsay, France and also with the Center for Computational Physics, Institute of Physics, VAST, P.O. Box 429 Bo Ho, Hanoi 10000, Vietnam (e-mail: viet-hung.nguyen@u-psud.fr).}\thanks{F. Mazzamuto, A. Bournel and P. Dollfus are with the Institut d'Electronique Fondamentale, CNRS, Universit$\acute{e}$ Paris-Sud, UMR8622, 91405 Orsay, France.}} 

\maketitle

\begin{abstract}
     In this letter, we propose the resonant tunneling diode (RTD) based on a double-barrier graphene/boron nitride (BN) heterostructure as device suitable to take advantage of the elaboration of atomic sheets containing different domains of BN and C phases within a hexagonal lattice. The device operation and performance are investigated by means of a self-consistent model within the non-equilibrium Green's function formalism on a tight-binding Hamiltonian. This RTD exhibits a negative differential conductance effect which involves the resonant tunneling through both the electron and hole bound states of the graphene quantum well. It is shown that the peak-to-valley ratio can reach the value of 4 at room temperature for gapless graphene and the value of 13 for a bandgap of 50 meV.
\end{abstract}

\begin{IEEEkeywords}
Graphene, Boron nitride substrate, Resonant tunneling diode.
\end{IEEEkeywords}

\section{Introduction}

\IEEEPARstart{T}{he} benefits of the extraordinary intrinsic transport properties of graphene \cite{bolo08} are usually hindered by the defects of the supporting insulating substrate. Indeed, graphene on various substrates such as SiO$_2$ \cite{chen08}, SiC \cite{lin010} or other high-$\kappa$ insulators \cite{pono09} shows a reduced carrier mobility, which is usually attributed to surface roughness, charge surface states or surface optical phonons. Recently, it has been shown that graphene reported on hexagonal boron nitride (h-BN) has higher mobility than on any other substrate \cite{dean10,jxue11,deck11}. A mobility of 275 000 cm$^2$/Vs at low temperature and 125 000 cm$^2$/Vs at room temperature, i.e. as high as for suspended graphene, has even been reported \cite{zome11}. It is due to the fact that the surface of h-BN is flat, with a low density of charged impurities, does not have dangling bonds and is relatively inert \cite{dean10}. Hence, with the same atomic structure as graphene and a 1.8$\%$ higher lattice constant \cite{liu003} h-BN is becoming a very promising candidate as high bandgap insulating material to make possible reaching the ballistic transport regime in graphene devices even at room temperature and exploiting the peculiarities of graphene properties inherent in the massless and chiral character of charge carriers. Additionally, it has been shown that it is possible to obtain atomic sheets containing different domains of h-BN and C phases with various compositions \cite{ci0010}, which offers the perspective of designing new heterostructures and devices \cite{fior11}.

In the field of quantum devices, graphene resonant tunnelling diodes (RTDs) have been proposed and investigated previously \cite{teon09,vhn011}. However, while the operation of graphene nanoribbon (GNR) RTDs is strongly dependent on the device shape \cite{teon09}, the 2D graphene structure studied in \cite{vhn011} assumes that a bandgap opens in graphene epitaxially grown on SiC, which is a debated question, in addition to the substrate-induced mobility reduction. In this work, we thus propose a new graphene RTD based on graphene/h-BN heterostructures, where the two barriers are assumed to be formed by large bandgap h-BN zones, as schematized in Fig. 1. Though it may be useful and possible in case of Bernal stacking on BN substrate \cite{khar11}, the bandgap in graphene is not necessary for this device to operate correctly.

To investigate the electrical characteristics of this RTD, we have developed a self-consistent simulation code based on the Green's function formalism \cite{vhn009} to solve the tight-binding Hamiltonian suggested in \cite{seol11}.
\begin{figure}[!t]
\centering
\includegraphics[width=3.1in]{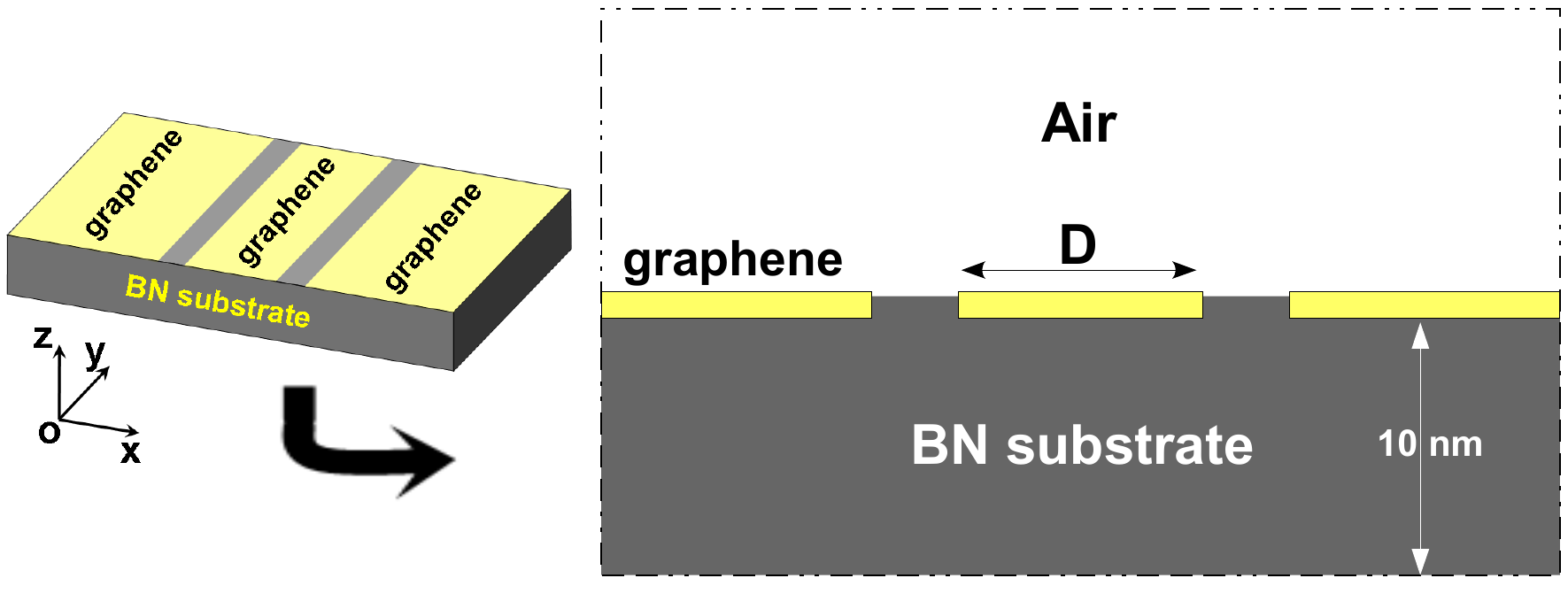}
\caption{Simulated device wherein the graphene sheets are deposited on a h-BN substrate and two barriers are thus formed by the BN layers between the graphene zones.}
\label{fig_sim0}
\end{figure}

\section{Physical model and results}

In the structure schematized in Fig. 1, the BN barriers are inserted in a monolayer graphene sheet reported on a h-BN substrate of thickness 10 nm and dielectric constant $\epsilon_r$ = 3.5 \cite{dean10}. We assume the transverse width along OY direction to be much larger than the typical device length along the transport direction, i.e. a few tens of nanometers, so that the extension along the transverse direction can be considered through Bloch periodic boundary conditions \cite{fior09}. The length of the heavily doped regions at both device ends and the BN barrier thickness are assumed to be 17 nm and 1.3 nm, respectively. Our approach is based on the self-consistent solution of the 2D Poisson and Schr\"{o}dinger equations. The nearest-neighbor tight-binding Hamiltonian with parameters determined in \cite{seol11} is solved using the Green's function method \cite{vhn009} in the ballistic approximation. In this simulation, the null Neumann conditions are applied at the 2D-domain boundaries (dashed line in Fig. 1). In the case of Bernal stacking, graphene reported on a h-BN substrate was shown to exhibit a finite bandgap, but this bandgap vanishes when graphene is misaligned with respect to h-BN lattice \cite{khar11}. Therefore, unless otherwise stated, the zero bandgap of graphene was assumed in our study. All simulations were performed at room temperature.
\begin{figure}[!t]
\centering
\includegraphics[width=3.2in]{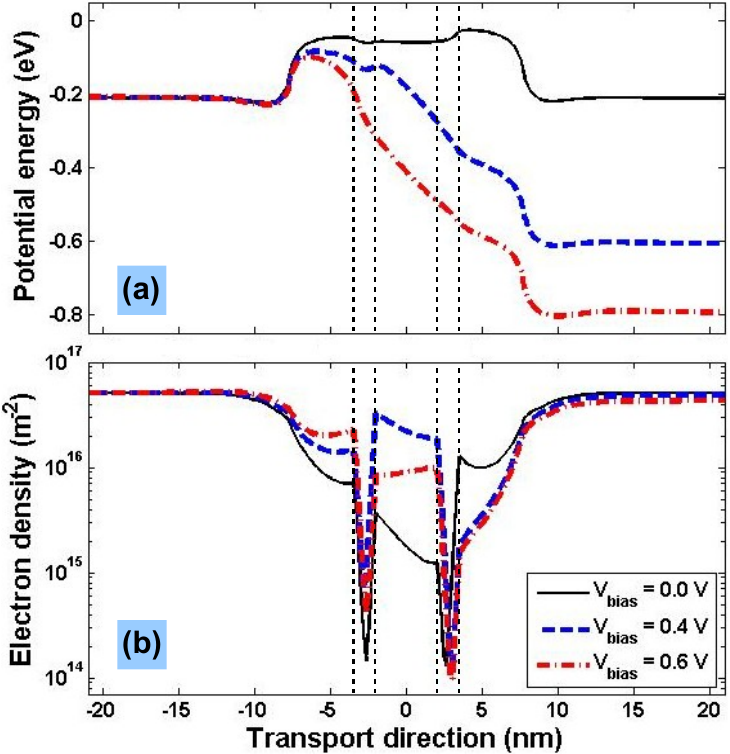}
\caption{(a) Self consistent data of potential energy and (b) corresponding electron density at different applied biases. The doped concentration in two device ends $N_D = 5 \times 10^{16} m^{-2}$. The vertical-dashed lines indicate the h-BN regions.}
\label{fig_sim1}
\end{figure}

\begin{figure*}[!t]
\centering
\includegraphics[width=6.0in]{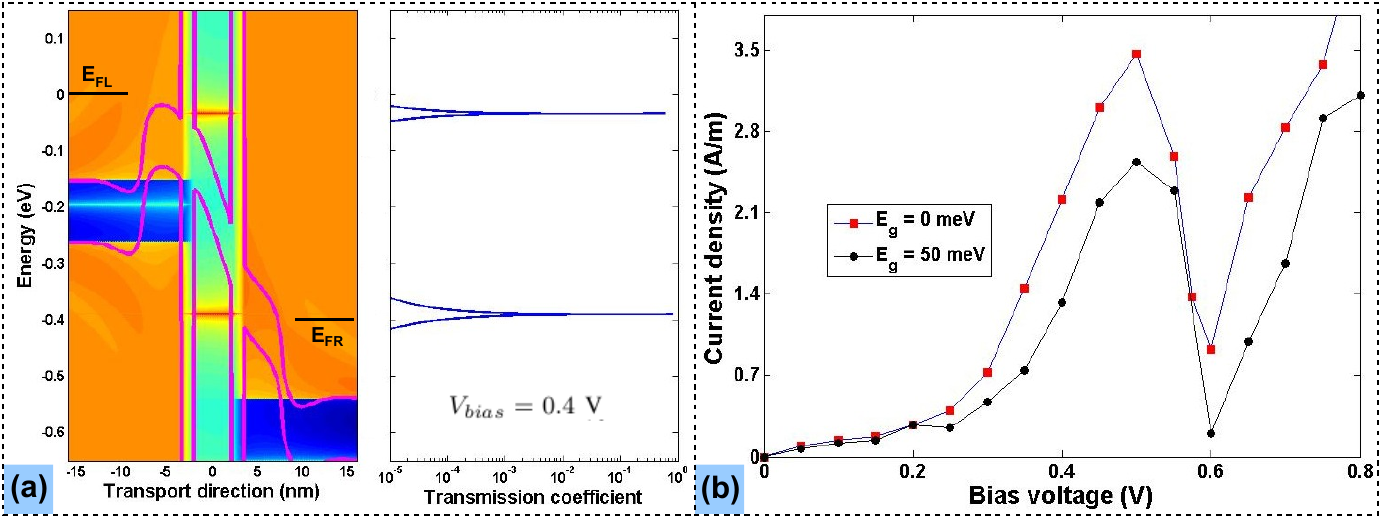}
\caption{(a) Diagram showing the LDOS and the transmission coefficient as a function of energy at $V_{bias} = 0.4$ V and (b) I-V characteristics of the simulated device (for both cases of zero and finite bandgaps). The figures in (a) are plotted for a transversal momentum $k_y \neq K_y$ (momentum at the Dirac point). The corresponding electron/hole potential profiles are superimposed on the LDOS (solid lines). The doped concentration $N_D = 5 \times 10^{16} m^{-2}$.}
\label{fig_sim2}
\end{figure*}
In Fig. 2, the self-consistent solutions of potential and electron density are plotted for different biases applied to the device with well width $D$ = 4.3 nm and doping concentration in access regions $N_D = 5\times10^{16}$ m$^{-2}$. The low doping in the central zone generates a potential barrier which controls the threshold voltage of the device. Differently from the conventional RTDs \cite{vndo06}, Fig. 2(b) shows that the local charge density is not symmetrical even at zero bias. This feature can be understood as a consequence of the device asymmetry between the left graphene section coupled to nitrogen atoms and the right section bounded to boron atoms of the h-BN lattice. Besides, it is shown that the electron density accumulated in the quantum well increases when increasing the bias voltage from 0 V to 0.4 V and decreases when further raising the bias from 0.4 V to 0.6 V. This feature is nothing but an effect of resonant tunneling \cite{vndo06} as described below.

We now go to explore the electrical behavior of the device. In Fig. 3, we display the I-V characteristics together with a diagram showing the LDOS and the corresponding transmission coefficient at $V_{bias}$ = 0.4 V. Indeed, the diagram in Fig. 3(a) shows that the large bandgap ($\approx 3.9$ eV \cite{seol11}) in the h-BN barriers generates the quantization of carrier states in the graphene quantum well, which gives rise to resonant tunneling effects. As a consequence, the I-V curves can exhibit a strong negative different conductance (NDC) in both cases of gapless and gapped graphene. In principle, the valley current in semiconductor RTDs occurs at a finite bias when there is no available state in the emitter region for tunneling via the confined states in the quantum well. This feature is essentially due to the finite bandgap of the emitter. The results obtained here are thus explained as follows. Though the energy bandgap of graphene may be actually zero for the transverse momentum mode corresponding to the Dirac (or K) point (i.e., $k_y \equiv K_y = 2\pi/3\sqrt{3}a_c$), a finite energy gap $\hat E_g (k_y) = 2t_{CC}\left| 1 - 2cos\left( a_ck_y\sqrt{3}/2 \right) \right|$ still appears for the other modes (see in ref. \cite{neto09}), where $t_{CC} = 2.5$ eV \cite{seol11} and $a_c$ is the carbon-carbon distance. For instance, $\hat E_g (k_y) \approx 0.11$ eV for $k_y  = K_y + 0.008\pi/a_c \sqrt 3$ as seen in the diagram of Fig. 3(a). Therefore, since the total current results from the contribution of many transverse momentum modes, the valley current occurs even in gapless graphene devices as seen in Fig. 3(b) when the tunneling corresponding to finite values of $\left| k_y - K_y \right|$ is suppressed at high bias. Moreover, because of the zero (or small) bandgap of graphene, it also shows that the current valley in this device is much narrower than that in conventional RTDs \cite{vndo06}. Besides, when introducing a finite bandgap \cite{teon09,khar11} in graphene, both the valley and peak currents are reduced while the peak-to-valley current ratio (PVR) increases. The latter reaches 4 and 13 for zero bandgap and $E_g$ = 50 meV, respectively.

\begin{figure}[!t]
\centering
\includegraphics[width=3.2in]{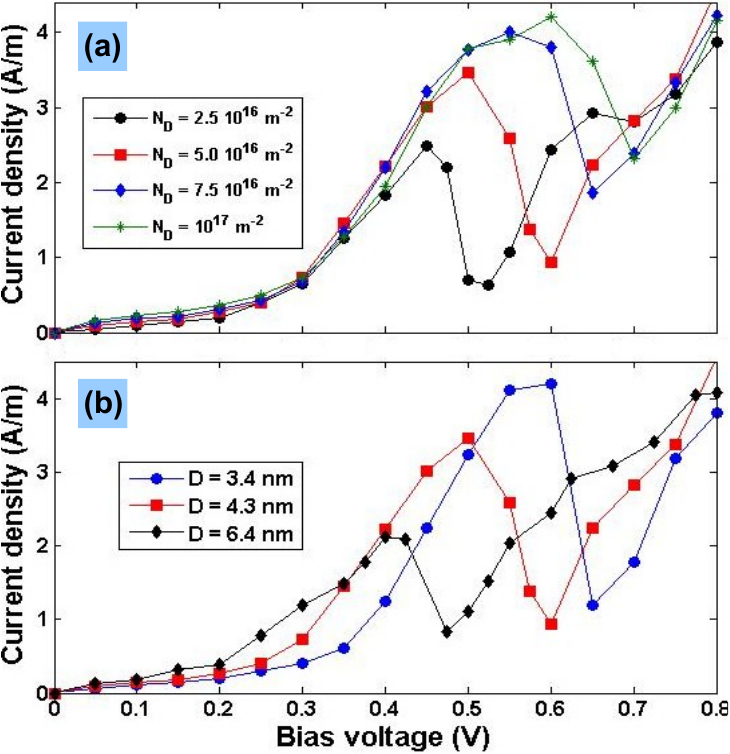}
\caption{I-V characteristics (a) for different doped concentrations in two contacts and (b) for different well widths $D$. $D = 4.3$ nm in (a) and $N_D = 5\times10^{16}$ m$^{-2}$ in (b).}
\label{fig_sim3}
\end{figure}
Especially, from Fig. 3, we find an interesting feature: differently from other RTDs \cite{teon09,vndo06} where only the electron states in the quantum well (QW) are resonant, in this device, both electron and hole states can contribute to the current, as a consequence of the chirality of carriers in graphene. This point has been also discussed in \cite{vndo08} and was shown to reduce the NDC effect in single potential barrier graphene structures. Practically, as we can understand from Fig. 3(a), the confined hole state in the QW makes the tunneling from hole states in the emitter to electron states in the collector possible. This band-to-band (BTB) process contributes to the rapid increase of current beyond $V_{bias}$= 0.25 V, in addition to the normal resonant tunneling of electrons via the electron state in the QW. The vanishing of the latter process is responsible for the valley regime in the I-V characteristics. When further increasing $V_{bias}$, the BTB tunneling through the first electron level in the QW occurs, which leads to the rapid re-increase of current and a narrow current valley seen in Fig. 3(b).

To evaluate the role of device parameters on the RTD operation, we display in Figs. 4(a) and (b) the I-V characteristics obtained for different doping concentrations $N_D$ in the access regions and for different well widths $D$, respectively. In fact, when increasing $N_D$, the energy spacing between the Fermi level (zero-energy point) and the flat potential energy in the device ends at zero bias increases. This results in shifting the valley region to high bias. Besides, the enhancement of BTB resonant tunneling at high bias makes both the peak and valley currents higher when increasing $N_D$. It finally reduces the PVR which reaches $\sim$ 3.9, 3.8, 2.2 and 1.8 for $N_D = 2.5\times10^{16}$, $5.0\times10^{16}$, $7.5\times10^{16}$, and $10^{17}$ $m^{-2}$, respectively. Fig. 4(b) shows that when increasing the well width $D$, the current valley moves to the low bias while the NDC effect generally weakens. This feature can be straightforwardly understood as the influence of $D$ on the position and number of energy levels in the QW. In the case of large $D$ (e.g., 6.4 nm in Fig. 4(b)), the multi-level contribution gives rise to a quite complex I-V curve with low PVR.

Finally, though not shown here, the overall current appears to be strongly reduced when increasing the BN barrier thickness, as a consequence of the large bandgap of h-BN. It therefore suggests that to observe well the effects discussed above, a barrier thickness smaller than 2 nm is mandatory.

\section{Conclusion}

A simulation study of graphene RTDs based on graphene/h-BN double-barrier structures has been performed by means of the self-consistent solution of the 2D Poisson and Schrödinger equations within the tight-binding. It was shown that the resonant tunneling and the resulting NDC behavior may involve both the electron and hole bound states of the graphene quantum well. The sensitivity of the electrical characteristics to the device parameters as the doping density of access zones, the QW thickness and the barrier thickness has been analyzed. Though the chiral band-to-band tunneling tends to reduce the width of the valley region in the I-V characteristics and to induce a rapid re-increase of current, the PVR of NDC effect can reach the value of 4 at room temperature for gapless graphene and the value of 13 for a small bandgap of 50 meV which may result from Bernal stacking of graphene on h-BN. This work suggests that the engineering of graphene/h-BN structures is opening a new route for high-performance graphene devices.

\section*{Acknowledgment}

This work was partially supported by the French ANR through the projects NANOSIM-GRAPHENE (ANR-09-NANO-016) and MIGRAQUEL (ANR-10-BLAN-0304).

\ifCLASSOPTIONcaptionsoff

\fi


\begin{thebibliography}{1}
\bibitem{bolo08} K. I. Bolotin \emph{et al.}, Solid State Comm. \textbf{146}, 351 (2008).
\bibitem{chen08} J.-H. Chen \emph{et al.}, Nat. Nanotechnol. \textbf{3}, 206 (2008).
\bibitem{lin010} Y.-M. Lin \emph{et al.}, Science \textbf{327}, 662 (2010).
\bibitem{pono09} L. A. Ponomarenko \emph{et al.}, Phys. Rev. Lett. \textbf{102}, 206603 (2009).
\bibitem{dean10} C. R. Dean \emph{et al.}, Nat. Nanotechnol. \textbf{5}, 722 (2010).
\bibitem{jxue11} J. Xue \emph{et al.}, Nat. Mater. \textbf{10}, 282 (2011).
\bibitem{deck11} R. Decker \emph{et al.}, Nano Lett. \textbf{11}, 2291 (2011).
\bibitem{zome11} P. J. Zomer \emph{et al.}, Appl. Phys. Lett. \textbf{99}, 232104 (2011).
\bibitem{liu003} L. Liu \emph{et al.}, Phys. Rev. B \textbf{68}, 104102 (2003).
\bibitem{ci0010} L. Ci \emph{et al.}, Nat. Mater. \textbf{9}, 430 (2010).
\bibitem{fior11} G. Fiori \emph{et al.}, IEDM Tech. Dig., pp. 259-262 (2011).
\bibitem{teon09} H. Teong \emph{et al.}, J. Appl. Phys. \textbf{105}, 084317 (2009).
\bibitem{vhn011} V. Hung Nguyen \emph{et al.}, Semicond. Sci. Technol. \textbf{26}, 125012 (2011).
\bibitem{khar11} N. Kharche \emph{et al.}, Nano Lett., DOI: 10.1021/nl202725w
\bibitem{vhn009} V. Hung Nguyen \emph{et al.}, J. Appl. Phys. \textbf{106}, 053710 (2009).
\bibitem{seol11} G. Seol \emph{et al.}, Appl. Phys. Lett. \textbf{98}, 143107 (2011).
\bibitem{fior09} G. Fiori \emph{et al.}, IEEE Electron Device Lett. \textbf{30}, 261 (2009).
\bibitem{vndo06} V. Nam Do \emph{et al.}, J. Appl. Phys. \textbf{100}, 093705 (2006).
\bibitem{neto09} A. H. Castro Neto \emph{et al.}, Rev. Mod. Phys. \textbf{81}, 109 (2009).
\bibitem{vndo08} V. Nam Do \emph{et al.}, J. Appl. Phys. \textbf{104}, 063708 (2008).
\end{thebibliography}
\end{document}